\documentclass[aps,pra,showpacs,showkeys,superscriptaddress,nofootinbib]{revtex4}

\newcommand{\n}{\nonumber}
\newcommand{\bra}[1]{\langle#1|}
\newcommand{\ket}[1]{|#1\rangle}

\usepackage{color}

\begin{document}

\title{Multi-qudit states generated by unitary braid quantum gates\\ based on
Temperley-Lieb algebra}

\author{C.-L. Ho
%\footnote{email: hcl@mail.tku.edu.tw\\
%Tel: +886-2-26215656 Ext 2521\\
%Fax: +886-2-26209917}
}
 \affiliation{Department of Physics, Faculty of Core Research,
Ochanomizu University\\
2-1-1 Ohtsuka, Bunkyo-ku, Tokyo 112-8610, Japan}
\affiliation{Department of Physics, Tamkang University, Tamsui
25137, Taiwan, R.O.C.\footnote{Permanent address}}

\author{T. Deguchi}
\affiliation{Department of Physics, Faculty of Core Research,
Ochanomizu University\\
2-1-1 Ohtsuka, Bunkyo-ku, Tokyo 112-8610, Japan}

%\date{28 Jun 2017} 14 Apr 2017}, % 5 Apr 2017}    %4 Jan 2017}
\begin{abstract}

Using a braid group representation based on the Temperley-Lieb algebra, we construct braid quantum gates that could generate entangled  $n$-partite $D$-level qudit states.
$D$ different sets of $D^n\times D^n$  unitary representation of the braid group generators are presented. 
With these generators the desired braid quantum gates are obtained. 
We show that the generalized GHZ states, which are maximally entangled states, can be obtained directly from these braid quantum gates without resorting to further local unitary transformations.
We also point out an interesting observation,  namely for a general multi-qudit state there exists a unitary braid quantum gate based on the Temperley-Lieb algebra that connects it from one of its component basis states, if the coefficient of the component state is such that the square of its norm is no less than $1/4$.

\end{abstract}

\pacs{03.67.Bg, 02.10.Kn, 03.67.Mn}
%(Entanglement production and manipulation, knot theory,
%entanglement measures, witnesses, and other characterizations)
%\keywords{Entanglement, Temperley-Lieb algebra, braid
%representation, GHZ states, cluster states}
\keywords{quantum entanglement, braid operators, Temperley-Lieb algebra }

\maketitle

%%%%%%%%%%%%%%%%%%%%%%%%%%
\section{Introduction}

Topological quantum computing
is an approach to fault-tolerant quantum computing based on topological phases of matter. 
Most models of this approach are based on the braiding and fusing of anyons,
 quasi-particles which obey fractional statistics in two dimensions \cite{K1,K2,FLZ,N}.  
  Interest in this approach is enhanced in recent years by the prospect that anyons 
  could be practically realized through Majorana zero modes \cite{M1,M2,M3} in topological insulators 
  \cite{FK,B}.

These interesting  developments in  quantum computation using
the braiding of anyons have
stimulated interest in applying the theory of braid groups to the
fields of quantum information and quantum computation. 
 In this
respect, an interesting result is the realisation that a specific
solution of the Yang-Baxter equation,  the so-called Bell basis change matrix, 
can be obtained from the braiding relations \cite{Kauff04}.
It is known that this operator
is  a universal gate for quantum computing in
the presence of local unitary transformations . 
As such, the result shown in \cite{Kauff04}  implies that in principle all quantum
gates can be constructed from braiding operators together with
single qubit gates.

The braid operator presented in \cite{Kauff04} involves a solution of the Yang-Baxter equation.
It generates the four
maximally entangled Bell states from the standard basis of
separable states. 
This has led  to further investigation on the
possibility of generating other entangled states  by appropriate
braiding operators \cite{ZKG,CXG07,ZJG,HSO,WSLLZX}.  
For instance, in \cite{CXG07} unitary
braiding operators based on the Yang-Baxter equation were used to realise entanglement swapping and
generate the Greenberger-Horne-Zeilinger (GHZ) state \cite{GHZ}  and the linear cluster state \cite{cluster}.
The GHZ state is of fundamental importance  as it is
 the maximally entangled multipartite state, which includes the Bell states as special cases.

In \cite{HSO} it was shown 
how the Bell
states, the generalized GHZ states, some cluster-like states, 
and other states with varying degrees of entanglement
may be generated {\em directly} from a braiding operator. 
Instead of using the Yang-Baxter equation, 
there we adopt
a different approach.  A new type of braiding operators were obtained by a unitary Jones representation of the braid group \cite{Jones85}, which is constructed using a new representation of the Temperley-Lieb algebra (TLA)
\cite{TL}.
This new class of
braiding operators 
generalizes the Bell matrix to multi-qubit systems, thus unifying
the Hadamard, Bell and GHZ matrices within the same framework.

In recent years there is also a growing interest in 
generalising the braiding operators 
 to  quantum systems with higher ($D$-level) dimensional Hilbert space, so-called qudits,
 both theoretical \cite{Cerf,ACDM,X1,X2,X3,X4,Zhou,D1,D2,D3,D4,D5,D6} and experimental \cite{Mar}.
 This is due partly to the potential in enhanced security in quantum cryptography 
 offered by qudit states as compared to qubit states \cite{Cerf}, and partly to the 
 potentially richer mathematical structures inherent in these systems.
 The Yang-Baxter approach was employed to construct braid operator for the bipartite qudit systems in \cite{ACDM,X1,X2,X3}.
  In \cite{Zhou} multi-qudit cluster state was considered in a non-braid group approach based on the quantum plane algebra \cite{Weyl,Sch}.   Recently, this latter algebra was also employed in \cite{X4} to obtain a representation of braid operator
 for the multi-qudit systems.  
  
In this work we would like to show how the braid group representation based on the Temperley-Lieb algebra considered in
\cite{HSO} for the qubit systems could be extended to multipartite qudit systems. 

We shall first briefly review in Sect.\,II the relation between quantum entanglement  and Jones representation of the braid group based on Temperley-Lieb algebra.    We then discuss the unitary Jones representation  of braid operators for the qubit systems in Sect.\,III, and extend it to qudit systems in Sect.\,IV.  Braid quantum gates that can entangle multi-qudit states, and the generalized GHZ states are 
discussed in Sect.\,V.  Sect.\,VI discusses how a general qudit state may be generated from one of its component state by a braid operator under appropriate condition. Sect.\,VII concludes the paper.

%%%%%%%%%%%%%%%%%%%%%%%%%%%%%%%%%%
\section{Braid group and quantum entanglement}
%%%%%%%%%%%%%%%%%%%%%%%%%%%%%%%%%%

Let us first briefly review the relation between braid group and quantum entanglement \cite{Kauff04,HSO}.

The $m$-stranded
braid group $B_m$ is generated by a set of elements
$\{\sigma_1,\sigma_2,\ldots,\sigma_{m-1}\}$ with  defining relations:
\begin{eqnarray}
\sigma_i \sigma_j &=& \sigma_j \sigma_i,~~|i-j|>1;\nonumber\\
\sigma_i \sigma_{i+1} \sigma_i &=& \sigma_{i+1} \sigma_i \sigma_{i+1},~~1\leq i <m. \label{BGR}
\end{eqnarray}
To apply the braid group in
quantum computing, one needs its unitary representations, because quantum gates are represented by
unitary operators. For
an $m$-qubit system the  $2^m\times2^m$ unitary
representation of $B_m$ commonly employed is
\begin{eqnarray}
\sigma_i= I \otimes\ldots \otimes I \otimes\ R \otimes I\otimes \ldots
\otimes I
\label{R} \; \; \; \; \; (i=1\ldots m-1)\, ,
\end{eqnarray}
where $I$ is the $2\times 2$ unit matrix and $R$ is a $4\times 4$
unitary matrix that acts on both the $i$-th and $(i+1)$-th qubits;
that is, occupying the $(i,i+1)$ position.  This choice of the $\sigma_i$'s satisfy the first of the two
braid group relations in (\ref{BGR}) automatically. To fulfil the second relation, $R$ must
satisfy
\begin{eqnarray}
\left(R\otimes I\right)\left(I\otimes R\right)\left(R\otimes
I\right)=\left(I\otimes R\right)\left(R\otimes
I\right)\left(I\otimes R\right).\label{YBE}
\end{eqnarray}
This relation is usually called the (algebraic) Yang-Baxter
equation. 

One of the simplest solutions of (\ref{YBE}) that
produces entanglement of states is the Bell matrix
\begin{eqnarray}
R= \frac{1}{\sqrt{2}}
 \left( \begin{array}{cccc}
  1 & 0 & 0 &-1 \\
  0 & 1 &-1 & 0 \\
  0 & 1 & 1 & 0 \\
  1 & 0 & 0 & 1
  \end{array}\right)\, .
  \label{Bell-Mat}
\end{eqnarray}
 When acting on the standard basis $\{|00\rangle, |01\rangle,
|10\rangle, |11\rangle\}$, $R$ generates the four maximally
entangled Bell states $(|00\rangle \pm |11\rangle)/\sqrt{2}$ and
$(|01\rangle \pm |10\rangle)/\sqrt{2}$. Here we adopt the
convention $|0\rangle=(1,0)^t$ and $|1\rangle=(0,1)^t$, where $t$
denotes the transpose.
In the presence of local unitary transformations, $R$ is a
universal gate \cite{Kauff04}.
The Bell matrix can be viewed as the bipartite generalisation of the Hadamard matrix
\begin{eqnarray}
H=\frac{1}{\sqrt{2}}\left(
\begin{array}{cc}
    1 & 1 \\
    1 & -1  \\
\end{array}\right),
\end{eqnarray}
which when acts on the single qubit $|0\rangle$ and $|1\rangle$ gives a linear superposition of them.

 In constructing
the well-known Jones polynomials for knots
and links, Jones \cite{Jones85} provided a new representation of
the braid group based on what is essentially the TLA. The TLA,
denoted by $TL_m(d)$, is defined, for an integer
$m$ and a complex number $d$, to be the algebra generated by the
unit element $I$ and the elements $h_1,h_2,\ldots,h_{m-1}$
satisfying the relations
\begin{eqnarray}
h_ih_j&=&h_jh_i,~~|i-j|>1;\nonumber\\
h_ih_{i\pm 1}h_i &=& h_i,~~1\leq i <m, \label{TLA}\\
h_i^2 &=& dh_i.\nonumber
\end{eqnarray}
The Jones representation of the braid group is then
given by (see eg., \cite{Kauff1})
\begin{eqnarray}
\sigma_i=Ah_i + A^{-1} I,~~ \sigma_i^{-1}=A^{-1}h_i + AI,\label{JonesRep}
\end{eqnarray}
where $A$ is a complex number given by $d=-A^2-A^{-2}$. 

Generally the Jones representation is  not unitary.  However, if $A=e^{i\theta}$
($\theta\in [0,2\pi)$) and all the $h_i$'s are Hermitian
($h_i^\dagger =h_i$), then the Jones representation is
unitary.  For $A=e^{i\theta}$, $d=-2\cos 2\,\theta$ is real. 
This also implies that $d^2\leq 4$.

Based on the unitary Jones representation of the 3-stranded braid group $B_3$, we have 
generalized the Hadamard and Bell matrices to higher
dimensions (i.e., to $n$ qubits), so that they generate
generalized GHZ states from separable states directly \cite{HSO}.  
In view of the recent interest in the qudit systems, here we would like to extend our construction 
in \cite{HSO} to the case of multi-qudit systems.

%%%%%%%%%%%%%%%%%%%%%%%%%%%%%%%%%%
\section{Unitary Jones representation of $B_3$: $2$-level  Qubits}
%%%%%%%%%%%%%%%%%%%%%%%%%%%%%%%%%%

The construction in \cite{HSO} is based on the following simple observation. 

For single qubit systems, 
 a simple
unitary Jones representation of $B_3$ can be given by the TLA elements $h_i=d E_i$ ($i=1,2$), where
\begin{eqnarray}
E_1 =\left( \begin{array}{cc} 1 & 0\\0 & 0
\end{array}\right),~~~
E_2 = \left( \begin{array}{cc} a^2 & e^{-i\phi} ab\\e^{i\phi} ab &
b^2
\end{array}\right),~~a^2+b^2=1.\label{E12}
\end{eqnarray}
Here $\phi$ is a phase angle. The $E_i$'s satisfy
\begin{eqnarray}
E_i^2 &=& E_i,\nonumber\\
E_1E_2E_1 &=& a^2 E_1,\label{E-alg}\\
E_2E_1E_2 &=& a^2 E_2.\nonumber
\end{eqnarray}
With $a^2=d^{-2}$, $h_i$'s as constructed from $E_i$'s satisfy the
TLA.   
For unitary representation of the braid group, one must have
$d^2\leq 4$, and so $a^2=d^{-2} \geq 1/4$.

Now as $d$ and $a$ are real, in order that $h_i$'s be
Hermitian, we must have $b^2=1-1/d^2 \geq 0$. This implies
$d^2\geq 1$ as well as the condition $d^2\leq 4$ mentioned before.  Hence $\theta\, ({\rm mod}\ 2\pi)$ is restricted to
be in the range $|\theta|\leq \pi/6$, $|\theta-\pi/2|\leq \pi/6$\footnote{We take this opportunity to add this range which was somehow neglected in \cite{HSO}.} or $|\theta-\pi|\leq \pi/6$.
We shall assume $\theta$ to be in these domains hereafter.  The
special case of this representation with $\phi=0$ was employed
previously in exploring the relation between quantum computing and
the Jones polynomials \cite{Kauff1,Kauff2}).

To generalise the above representation of TLA to
higher dimensions, we consider the following elements
\begin{eqnarray}
e_1= \left( \begin{array}{cc} 1 & 0\\0 & 0
\end{array}\right),\,e_2=\left( \begin{array}{cc} a^2 & 0\\ 0 & b^2
\end{array}\right),\,e_3=\left( \begin{array}{cc}  0 & e^{-i\phi} ab\\ e^{i\phi} ab&
0\end{array}\right),~~~~a^2+b^2=1.
\end{eqnarray}
Here $\phi$ is a phase angle. 
Define
\begin{eqnarray}
E_1&\equiv& \otimes_{j=1}^{k-1}I \otimes e_1
\otimes_{j=k+1}^n I,\nonumber\\
E_2 &\equiv& \otimes_{j=1}^{k-1}I \otimes e_2
\otimes_{j=k+1}^n I \label{E-2}\\
&+&  \otimes_{j=1}^{k-1} \lambda_j \otimes e_3 \otimes_{j=k+1}^n
\lambda_j,\nonumber
\end{eqnarray}
where $\otimes_{j=1}^{m} \lambda_j =\lambda_1\otimes \lambda_2 \otimes \cdots
\otimes \lambda_m$. Here $\lambda_j$ is any Hermitian operator satisfying
$\lambda_j^2=I$ (this requirement is to make $E_2^2=E_2$, which will be shown in the next paragraph). For example, $\lambda_j$ can be $I$, any one of the Pauli
matrices $\sigma_m (m=1,2,3)$, or the Hadamard matrix $H$. The
integer $n$ is the number of $2\times 2$ matrices in the tensor
products, and $k$ indicates the position of $e_1,~e_2$ and $e_3$.
The $E_i$'s are $2^n\times 2^n$ matrices, and they reduce
to (\ref{E12}) in the case of $n=k=1$. 
This freedom of choice in the $\lambda_j$'s in our approach allows us to construct braiding operators that could generate different degrees of qubit entanglement.

The crucial point of the representation (\ref{E-2}) is that the operators $e_1, e_2$ and $e_3$ satisfy the following identities:
\begin{eqnarray}
e_1^2 &=& e_1,\n\\
e_2^2+ e_3^2&=&e_2,\nonumber\\
e_2 e_3+ e_3 e_2 &=&e_3,\nonumber\\
e_1 e_2 e_1 &=& a^2 e_1,\nonumber\\
e_1 e_3 e_1 &=& 0,\label{Id}\\
e_2 e_1 e_2  &+ &e_3 e_1 e_3=a^2 e_2,\nonumber\\
e_2 e_1 e_3  &+& e_3 e_1 e_2=a^2 e_3.\nonumber
\end{eqnarray}
 Using these identities, one can easily check that
$E_i$'s satisfy (\ref{E-alg}). 
%---
For instance, we can check if $E_2^2=E_2$ is satisfied. From Eq.(\ref{E-2}) we have
\begin{eqnarray}
E_2^2 &=& \otimes_{j=1}^{k-1}I \otimes e_2^2
\otimes_{j=k+1}^n I +   \otimes_{j=1}^{k-1} \lambda_j^2 \otimes e_3^2 \otimes_{j=k+1}^n \lambda_j^2 \nonumber\\
&+&  \otimes_{j=1}^{k-1} \lambda_j \otimes \left(e_2 e_3 + e_3 e_2\right) \otimes_{j=k+1}^n
\lambda_j\, .
\end{eqnarray}
With $e_2^2+ e_3^2=e_2, 
e_2 e_3+ e_3 e_2 =e_3$, and $\lambda_j^2=I$, the equality $E_2^2=E_2$ is satisfied.

%--

Hence, the operators
$h_i=d E_i$ form a $2^n\times 2^n$ matrix
realization of $TL_3 (d)$.  A unitary braid group representation is then obtained
from the $h_i$'s by the Jones representation.
This new unitary braid representation generalizes the $2\times 2$
matrices of (\ref{E12}) to $2^n\times 2^n$ matrices of (\ref{E-2})
within the TLA $TL_3 (d)$. 

In the next section, we shall generalize the above construction to multipartite qudit systems.

%%%%%%%%%%%%%%%%%%%%%%%%%%%%%%%%%%%%
\section{Unitary Jones representation of $B_3$: $D$-level  Qudits }
%%%%%%%%%%%%%%%%%%%%%%%%%%%%%%%%%%%%

A  $D$-level qudit state is denoted by $\ket{s}$, where $s=0, 1,\ldots, D-1$, according to the convention that
$\ket{j}=(0,0,\ldots, 1, 0,\ldots)^t$ has an entry ``$1$" in the $(j+1)$ -th row, and ``$0$" elsewhere.
 The orthonormal computational basis of an $n$-qudit system is $\{\ket{s_1 s_2\cdots s_n}\}$.

To extend the previous construction to the multipartite $D$-level qudits, we  first note that  $e_1, e_2$ and $e_3$ can be expressed as
\begin{eqnarray}
e_1=\ket{0}\bra{0}, ~e_2=a^2\ket{0}\bra{0}+b^2 \ket{1}\bra{1},~~ 
e_3=ab\left(e^{-i\varphi}\ket{0}\bra{1}+e^{i\varphi}\ket{1}\bra{0}\right).
\end{eqnarray}
This inspires us to consider replacing the qubit $\ket{1}$  by the qudit $\ket{l}, (l=1,2,\ldots, D-1)$ in the definition of $e_2$ and $e_3$.
Doing this we get 
$D-1$ sets of $e_i$'s:
\begin{eqnarray}
e_1 &=&\ket{0}\bra{0}, ~~~ e_2^{(l)}=a_1^2\ket{0}\bra{0}+b_1^2 \ket{l}\bra{l},~~ \n\\
e_3^{(l)} &=& a_l b_l\left(e^{-i\varphi_l}\ket{0}\bra{l}+e^{i\varphi_l}\ket{l}\bra{0}\right),
\quad\quad l=1,2,\ldots,D-1.
\label{e}
\end{eqnarray}
Here $a_l, b_l (l=1,2)$ are real constants satisfying $a_l^2+b_l^2=1$, and $\varphi_{l}$ are arbitrary phases.
The operator $e_1$ being the same for  all the different sets.

It is easy to demonstrate that the identities in Eq.\,(\ref{Id}) are satisfied by the above sets of operators separately for each $l$. 

With these sets of operators we can construct, for each value of $l$, the two corresponding TLA's generators as follows:
\begin{eqnarray}
E_1^{(l)}&\equiv& \otimes_{j=1}^{k-1}I \otimes e_1
\otimes_{j=k+1}^n I,\n\\
E_2^{(l)} &\equiv& \otimes_{j=1}^{k-1}I \otimes e_2^{(l)}
\otimes_{j=k+1}^n I \label{E-d}\\
&+&  \otimes_{j=1}^{k-1} \lambda_j \otimes e_3^{(l)} \otimes_{j=k+1}^n
\lambda_j^{(l)}.\n
\end{eqnarray}
Here $\lambda_j^{(l)}$ is any Hermitian operator satisfying $\lambda_j^{(l)2}=I$. We can choose either $\lambda_j^{(l)}=I$, or
\begin{eqnarray}
\lambda_j^{(l)}\equiv \ket{s_j}\bra{\tilde {s}_j^{(l)}}+ \ket{\tilde{s}_j^{(l)}}\bra{s_j} +\sum_{s_j^\prime\neq  s_j, \tilde{s}_j^{(l)}} \ket{s_j^\prime}\bra{s_j^\prime},
\end{eqnarray}
with the action $\lambda_j^{(l)} \ket{s_j}=\ket{\tilde{s}_j^{(l)}}$, 
$ \lambda_j^{(l)} \ket{\tilde{s}_j^{(l)}} =\ket{s_j}$, and other qudits $\ket{s_j^\prime}$ ($s_j^\prime\neq  s_j, \tilde{s}_j^{(l)}$) unchanged.

These operators satisfy the $TL_3(d) $ algebra with the constant $d$ given by $d_l^2=a_l^{-2}$ for each $l$.
The corresponding braid operators are then given by the Jones representation
\begin{eqnarray}
\sigma_i^{(l)}=A_l d_l E^{(l)}_i + A_l^{-1} I,~~ \sigma_i^{(l)-1}=A_l^{-1} d_l E^{(l)}_i + A_l I,\label{JonesRep}
\end{eqnarray}
where $A_l$ are complex numbers determined by $d_l=-A_l^2-A_l^{-2}$.
As before, $A_l$ being a pure phase requires $d_l^2\leq 4$, and $b^2_l=1-a_l^2=1-1/d_l^2\geq 0$ implies 
$d_l^2\geq 1$, i.e.,
$1\leq d_l^2\leq 4$, or $1/4\leq a^2_l \leq 1$.

For each $l$, these $\sigma_i^{(l)}$ ($i=1,2$) furnish a $D^n\times D^n$   unitary representation of the braid group.

%%%%%%%%%%%%%%%%%%%
\section{Braid quantum gates}
%%%%%%%%%%%%%%%%%%%%%

 We define the unitary braiding transformation representing
the action of the braid $\sigma_1\sigma_2$. This braiding operator is
evaluated to be
\begin{eqnarray}
&& B_{0l}(n,k)\equiv \sigma_1^{(l)}\sigma_2^{(l)}\nonumber\\
 &=& \otimes_{j=1}^{k-1}I
\otimes
 \left( d_l a_l^2 \ket{0}\bra{0}+ \left( d_l b_l^2 + A_l^{-2}\right)\ket{l}\bra{l}+ \sum_{m\neq 0,l}\, A_l^{-2} \ket{m}\bra{m}\right)
\otimes_{j=k+1}^n I\label{B0l}\\
&+& d_l a_l b_l\otimes_{j=1}^{k-1} \lambda_j^{(l)} \otimes
 \left( e^{i\varphi_l} \ket{l}\bra{0} - e^{-i\varphi_l}A_l^4 \ket{0}\bra{l}\right) \otimes_{j=k+1}^n
\lambda_j^{(l)}\,.\nonumber
\end{eqnarray}
Here we have explicitly included the pair $(n,k)$ to indicate that this operator acts on an $n$-qudit system, with the $k^{th}$ qudit being the ``pivotal" qudit --- linear superposition of states is effected at this position, while qudits at other positions are only modified by the $\lambda_j^{(l)}$'s.

The action of $B_{0l}(n,k)$ on the separable $n$-qudit state $|s_1s_2\cdots s_n\rangle$ ($s_j=0,1,\ldots, D-1$) is
\begin{eqnarray}
B_{0l}(n,k)\,\ket{s_1s_2\cdots s_{k-1}, 0, s_{k+1}\cdots s_n}
&=&(d_la_l^2) \ket{s_1s_2\cdots s_{k-1} ,0 ,s_{k+1}\cdots s_n} \n\\
&&~~~~~~~~~ + (e^{i\varphi_l}d_l a_l b_l ) \ket{\tilde{s}_1^{(l)}\tilde{s}_2^{(l)}\cdots \tilde{s}_{k-1}^{(l)}, l ,\tilde{s}_{k+1}^{(l)}\cdots
\tilde{s}_n^{(l)}},\n\\
B_{0l}(n,k)\,\ket{s_1s_2\cdots s_{k-1},l , s_{k+1}\cdots s_n}
&=&(d_l b_l^2 + A_l^{-2}) \ket{s_1s_2\cdots s_{k-1},l, s_{k+1}\cdots s_n} \n\\
&&~~~~~~~~~ + (e^{-i\varphi_l} A_l^4 d_l a_l b_l ) \ket{\tilde{s}_1^{(l)}\tilde{s}_2^{(l)}\cdots \tilde{s}_{k-1}^{(l)}, 0,\tilde{s}_{k+1}^{(l)}\cdots
\tilde{s}_n^{(l)}},  
\label{B-act}\\
B_{0l}(n,k)\,\ket{s_1s_2\cdots s_{k-1}, m, s_{k+1}\cdots s_n}
&=&A_l^{-2} \ket{s_1s_2\cdots s_{k-1} ,m , s_{k+1}\cdots s_n}, ~~~~~~
m\neq 0,\,l.\n
\end{eqnarray}
Here $|\tilde{s}_j^{(l)}\rangle\equiv \lambda_j^{(l)} |s_j\rangle$
($j=1,\ldots,k-1,k+1,\ldots,n$). 
Thus under the action of
$B_{0l}(n,k)$, the separable $n$-qudit state
$\ket{s_1s_2\cdots  0 \cdots s_n}$ is superimposed on the
state $\ket{\tilde{s}_1\tilde{s}_2\cdots l \cdots
\tilde{s}_n}$, and likewise 
$\ket{s_1s_2\cdots , l, \cdots s_n}$ is superimposed on the
state $\ket{\tilde{s}_0\tilde{s}_2\cdots,0, \cdots
\tilde{s}_n}$.
We have used the subscript $\{0l\}$ in $B_{0l}(n,k)$ to indicate this fact.
The state in (\ref{B-act})  is normalized, as $ (d_la_l^2)^2 + |e^{i\phi_l}d_l a_l b_l|^2 =1
$. Depending on the choice of the set of $\lambda_j^{(l)}$'s,
the resulting state will have
varying degree of entanglement. In particular, if all $\lambda_j^{(l)}=I$,
then the resulting state is separable, and
$B_{0l}(n,k)$ is simply a local unitary transformation.

% --------------------------------------------------------------- 
\subsection{Successive superpositions of states}
%---------------------------------------------------------------

We now consider successive actions of braid operators $B_{0l}(n,k)$ ($l=1,2,\ldots, D-1$) on the separable $n$-qudit 
$\ket{\phi_0} \equiv |s_1s_2\cdots s_{k-1} ,0, s_{k+1}\cdots s_n\rangle$. 

Clearly we have from Eq.~(\ref{B-act})
\begin{eqnarray}
\ket{\phi_1}\equiv B_{01}(n,k)\ket{\phi_0}
=(d_1a_1^2)\ket{s_1s_2\cdots s_{k-1} 0 s_{k+1}\cdots s_n}
+(e^{i\varphi_1}d_1 a_1b_1 )\ket{\tilde{s}_1^{(1)}\tilde{s}_2^{(1)}\cdots \tilde{s}_{k-1}^{(1)} 1 \tilde{s}_{k+1}^{(1)}\cdots
\tilde{s}_n^{(1)}}.
\end{eqnarray}
Acting $B_{02}(n,k)$ on the state $\ket{\phi_1}$ gives
\begin{eqnarray}
\ket{\phi_2} \equiv B_{02}(n,k)B_{01}(n,k)\ket{\phi_0}
=
&~~~~~~(d_2 a_2^2)(d_1a_1^2)\,\ket{s_1s_2\cdots s_{k-1} 0 s_{k+1}\cdots s_n} & \n \\
&~+ A_2^{-2}(e^{i\varphi_1}d_1 a_1b_1 ) \,\ket{\tilde{s}_1^{(1)}\tilde{s}_2^{(1)}\cdots \tilde{s}_{k-1}^{(1)} 1 \tilde{s}_{k+1}^{(1)}\cdots \tilde{s}_n^{(1)}} &\\
&+(e^{i\varphi_2}d_2 a_2 b_2 )(d_1 a_1^2) \, \ket{\tilde{s}_1^{(2)}\tilde{s}_2^{(2)}\cdots \tilde{s}_{k-1}^{(2)} 2 \tilde{s}_{k+1}^{(2)}\cdots
\tilde{s}_n^{(2)}}.& \n
\end{eqnarray}
It is  clear that the successive actions of $B_{0l}(n,k)$'s are to linearly superimpose the states
$\ket{s_1s_2\ldots s_{k-1} 0 s_{k+1}\cdots s_n}, 
\ket{\tilde{s}_1^{(1)}\tilde{s}_2^{(1)}\cdots \tilde{s}_{k-1}^{(1)} 1 \tilde{s}_{k+1}^{(1)}\cdots \tilde{s}_n^{(1)}},
\ldots,
\ket{\tilde{s}_1^{(1)}\tilde{s}_2^{(l)}\cdots \tilde{s}_{k-1}^{(l)} l \tilde{s}_{k+1}^{(l)}\cdots \tilde{s}_n^{(l)}}
$ ($l=1,2,\ldots, D-1$) successively. Hence
\begin{eqnarray}
\ket{\phi_l} &\equiv& B_{0l}(n,k)B_{0,l-1}(n,k)\cdots B_{02}(n,k)  B_{01}(n,k)\ket{\phi_0}\n\\
&=& \alpha_{l0}\ket{s_1s_2\cdots s_{k-1} 0 s_{k+1}\cdots s_n}
+ \sum_{p=1}^l \alpha_{lp} \ket{\tilde{s}_1^{(p)}\tilde{s}_2^{(p)}\cdots \tilde{s}_{k-1}^{(p)} p \tilde{s}_{k+1}^{(p)}\cdots
\tilde{s}_n^{(j)}},
 \label{phi_l}\\
&&  ~~~~~~~~\alpha_{lr}: {\rm\ complex\  constants}; ~~ l=1,2,\ldots, D-1;~~ r=0,1,\ldots,l.\n
\end{eqnarray}

Depending on the choice of the set of operators $\lambda_j^{(l)}$'s, the states $\ket{\phi_l}$ could be separable (e.g., if all $\lambda_j^{(l)}=I$), partially entangled, or completely entangled.  
Also, as $B_{0l}(n,k)$ is unitary, normalizability is maintained at each transformation. This can be checked explicitly. 
Acting $B_{0,l+1}(n,k)$ on $\ket{\phi_l}$, we get
\begin{eqnarray}
\ket{\phi_{l+1}} &\equiv& B_{0,l+1}(n,k)\ket{\phi_l}\n\\
&=& (d_{l+1} a_{l+1}^2 \alpha_{l0})\ket{s_1s_2\cdots s_{k-1}, 0 ,s_{k+1}\cdots s_n}\n\\
&+& \sum_{p=1}^l (A_{l+1}^{-2}\,\alpha_{lp}) \ket{\tilde{s}_1^{(p)}\tilde{s}_2^{(p)}\cdots \tilde{s}_{k-1}^{(p)}, p, \tilde{s}_{k+1}^{(p)}\cdots \tilde{s}_n^{(p)}} \label{l+1}\\
&+&  (e^{i\varphi_{l+1}}d_{l+1} a_{l+1} b_{l+1} \alpha_{l0})\, 
\ket{\tilde{s}_1^{(l+1)}\tilde{s}_2^{(l+1)}\cdots \tilde{s}_{k-1}^{(l+1)}, l+1, \tilde{s}_{k+1}^{(l+1)}\cdots
\tilde{s}_n^{(l+1)}}. \n
\end{eqnarray}
Using the relations $a_{l+1}^2+b_{l+1}^2=1$ and $d_{l+1}^2 a_{l=1}^2=1$, one can check that
\begin{eqnarray}
\langle \phi_{l+1} | \phi_{l+1}\rangle =\sum_{s=0}^{l} |\alpha_{ls}|^2 =\langle \phi_{l} | \phi_{l}\rangle.
\end{eqnarray}
So $\ket{\phi_{l+1}}$ is normalized if $\ket{\phi_l}$ is. It is obvious that for $l=1$, the state $\ket{\phi_1}$ is normalized.
Hence by induction the states $\ket{\phi_l}$ generated by the braid quantum gates $B_{0l}(n,k)$'s are normalized. 

%---------------------------------------------------------------
\subsection{Coefficients of $\ket{\phi_l}$}
%---------------------------------------------------------------

Eq.\,(\ref{l+1}) relates the coefficients of $\ket{\phi_{l+1}}$ to those of $\ket{\phi_l}$.  
 In this and the next subsection, for simplicity of notation, we shall take $d_l=1/a_l$ (i.e., we take the positive sign of the relation $d^2_l=a_l^{-2}$), and also set all the phase angles to zero, $\varphi_l=0$.  Then Eq.\,(\ref{l+1}) gives
\begin{eqnarray}
\alpha_{l+1,0} &=& a_{l+1}\,\alpha_{l0};\n\\
\alpha_{l+1,p} &=& A_{l+1}^{-2}\,\alpha_{lp}, ~~~  p=1,2,\ldots,l;\label{coeff}\\
\alpha_{l+1,l+1} &=& b_{l+1}\,\alpha_{l0}.\n
\end{eqnarray}

For the initial separable state $\ket{s_1s_2\cdots s_k,0, s_{k=1}\cdots s_n}$, the initial value of the coefficient is $\alpha_{00}=1$.
Then from Eq.\,(\ref{coeff}) we have the solution of the coefficients $\alpha_{0s}$ ($s=0,1,\ldots,\,l $):
\begin{eqnarray}
\alpha_{l0}  &=& a_l\, a_{l-1}\,\cdots \,a _2\, a_1,\n\\
\alpha_{l1}  &=& A_l^{-2} A_{l-1}^{-2} \cdots A_3^{-2}  A_2^{-2}\, b_1,\n\\
\alpha_{l2}  &=& A_l^{-2} A_{l-1}^{-2} \cdots A_3^{-2}\,  b_2\, a_1,\n\\
\alpha_{l3}  &=& A_l^{-2} A_{l-1}^{-2} \cdots A_4^{-2} \, b_3 \,a_2\, a_1,\label{coeff2}\\
&& \vdots\n\\
\alpha_{ll}  &=& b_l\, a_{l-1}\,\cdots\, a _2\, a_1.\n
\end{eqnarray}

%--------------------------------------------------------------- 
\subsection{Generalized GHZ qudit states}
%---------------------------------------------------------------

We define the $l$-level generalized GHZ  $n$-qudit state as
\begin{eqnarray}
\ket{GHZ}_l\equiv \sum_{r=0}^l \alpha_{lr}\, \ket{rrr\cdots rrr}, ~~ l=1,\ldots, D-1,
\label{GHZ_l}
\end{eqnarray}
for which all the component states have the same probability to be measured, i.e., with $|\alpha_{lr}|^2=1/(l+1)$ for all  $r$.  These are the maximally entangled multi-qudit states.

We show here that by appropriately choosing the initial state $\ket{\phi_0}$, the set of operators $\lambda_j^{(p)}$'s, and the sets of scalars $\{a_p, b_p\}$ ($p=1,\ldots, l$), one can make $\ket{\phi_l}$ in Eq.\,(\ref{phi_l}) a GHZ state.

We start with the initial state $\ket{\phi_0}=\ket{00\cdots 00}$, i.e., choose all $\ket{s_j}=\ket{0},~ (j=1,2,\ldots, k-1, k+1, \ldots, n)$.   Then select the $\lambda_j^{(p)}$ to be
\begin{eqnarray}
\lambda_j^{(p)}=\ket{0}\bra{p} + \ket{p}\bra{0} + \sum_{q\neq 0,p} \ket{q}\bra{q},~~~p=1,\ldots, l.
\end{eqnarray}
These $\lambda_j^{(p)}$ change the $j$-qudit $\ket{0}_j$ in $\ket{\phi_0}$ to $\ket{p}_j$, $\lambda_j^{(p)}\ket{0}_j=\ket{p}_j$.

Next we have to solve the constraints $|\alpha_{lr}|^2=1/(l+1)$ for the required sets of $\{a_p, b_p\}$. From Eq.\, (\ref{coeff2})  it is found that the solution is given by 
\begin{eqnarray}
& a_1^2 = \frac{l}{l+1},~~~ &b_1^2 =\frac{1}{l+1},\n\\
& a_2^2 = \frac{l-1}{l},~~~  & b_2^2  =\frac{1}{l},\n\\
&  ~~~\vdots &\n\\
& a_j^2 = \frac{l-j+1}{l-j+2},~~~ & b_j^2 =\frac{1}{l-j+2},\\
&  ~~~\vdots &\n\\
& a_l^2 =\frac{1}{2},~~~& b_l^2 =\frac{1}{2}.\n
\end{eqnarray}
To fulfil the conditions $1\leq d_j=1/a_j\leq 2$, we must take the positive roots for $a_j$, 
i.e., $1/\sqrt{2}\leq a_j=\sqrt{(l-j+1)/(l-j+2)}<1$. The sign of $b_j$ can be chosen differently for different GHZ states.
The phase $A_l^{-2}$ is solved from $d_l=-A_l^2 - A_l^{-2}=1/a_l$.

With these parameters, one has from Eq.\,(\ref{phi_l}) 
\begin{equation}
\ket{GHZ}_l=B_{0l}(n,k)B_{0,l-1}(n,k)\cdots B_{02}(n,k)  B_{01}(n,k)\ket{00\cdots 00}.
\end{equation}

 Hence the generalised $l$-level GHZ $n$-qudit state can be generated from the original state $\ket{00\cdots 00}$.

%%%%%%%%%%%%%%%%%%%%%%%%%%%%%%%%%%%%%
\section{Generation of a general state from one of its component state}
%%%%%%%%%%%%%%%%%%%%%%%%%%%%%%%%%%%%%

In this section we demonstrate that there exists a  TLA-based unitary braid quantum gate which connects a general multi-qudit state and one of its component basis states, if the coefficient of the component state is such that the square of its norm is no less than $1/4$.

For clarity of presentation in this section,  we  shall replace the notation $\ket{s_1s_2s_3\cdots}$  for the orthonormal computational basis of $n$-qudits  by $\{ \ket{ijk\cdots}_n\}$ ($i,j,k,\ldots=0,1,\ldots, D-1$).

Now consider a general normalized state $\ket{\psi}=\sum_{i,j,k,\cdots=0}^{D-1}\, \alpha_{ijk\cdots}\,\ket{ijk\cdots}$.  Let $\ket{\psi}$ contains a component state, $\ket{abc\cdots}$, say, i.e., $\ket{\psi}=\cdots + \alpha_{abc\cdots}\,\ket{abc\cdots}+ \cdots$.
Next we construct the following two projection operators from $\ket{abc\cdots}$ and $\ket{\psi}$,
\begin{eqnarray}
E_1 &=& \ket{abc\cdots}\bra{abc\cdots},\n\\
E_2 &=& \ket{\psi}\bra{\psi}. 
\label{E_G}
\end{eqnarray}
By construction $E_1$ and $E_2$ are Hermitian, and satisfy the first  set of equations of TLA in Eq.\,(\ref{E-alg}). It is straightforward to show that the second and the third set of  Eq.\,(\ref{E-alg}) are also satisfied,
\begin{eqnarray}
E_1 E_2 E_1 & =& |\alpha_{abc\cdots}|^2 E_1,\n\\
E_2 E_1 E_2 & =& |\alpha_{abc\cdots}|^2 E_2,
\end{eqnarray}
obviously with $|\alpha_{abc\cdots}|^{2}$  playing the role of the scalar $d^{-2}$ of the TLA,  $d^2=|\alpha_{abc\cdots}|^{-2}$. 
Now in order for the TLA constructed from $E_1$ and $E_2$ to be unitary, one must have $1\leq d^2\leq 4$. This requires
$1/4 \leq  |\alpha_{abc\cdots}|^2 \leq 1$. The r.h.s. of the constraint is always satisfied as $|\alpha_{abc\cdots}|^2\leq 1$ for a normalized $\ket{\psi}$.

We can then form the braid generators $\sigma_1$ and $\sigma_2$ by the Jone representation as discussed before. Applying the braid operator $\sigma_1 \sigma_2$ on the separable state $\ket{abc\cdots}$, we get
\begin{eqnarray}
\sigma_1\sigma_2\ket{abc\cdots}=d\alpha_{abc\cdots}^*\ket{\psi} +\left(d+ A^{-2}+ (d^2 |\alpha_{abc\cdots}|^2) A^2 \right)\ket{abc\cdots}.
\end{eqnarray}
Since $d^2 |\alpha_{abc\cdots}|^2=1$,  $d=-A^2-A^{-2}$ and $d\alpha_{abc\cdots}^*=\exp(i\arg\,\alpha_{abc\cdots})$, we finally obtain,
\begin{eqnarray}
\sigma_1 \sigma_2 \ket{abc\cdots}=\exp(i\arg\,\alpha_{abc\cdots})\,\ket{\psi}.
\end{eqnarray}
So the braid quantum gate generates the state $\ket{\psi}$, up to a pure phase, from its component state $\ket{abc\cdots}$.  Of course, one can always factor out the phase as a global phase from the definition of $\ket{\psi}$.  In this case, $\alpha_{abc\cdots}$ is a real constant, and  we will have $\sigma_1 \sigma_2 \ket{abc\cdots}=\ket{\psi}$.

For the $l$-level generalized GHZ state defined in Eq.\,(\ref{GHZ_l}), the coefficients all give equal value, 
$ |\alpha_{lr}|^2=1/(l+1)$. For  $|\alpha_{lr}|^2=1/(l+1)\geq 1/4$, one must have $l+1=D\leq 4$.  Thus GHZ state can be generated from its component state by the method presented here only for qudits with level $D\leq 4$.

%\end{eqnarray}

%%%%%%%%%%
\section{Summary and discussion}
%%%%%%%%%%

In this paper we have constructed  braid operators  that could generate entangled  $n$-partite $D$-level qudit states, using a  braid group representation based on the Temperley-Lieb algebra.  $D$ different sets of $D^n\times D^n$  unitary representation $\{\sigma_1^{(l)}, \sigma_2^{(l)}\}$ ($l=1,2,\ldots,D$) of the braid generators were presented. 
Using these generators the desired braid quantum gates $B_{0l}(n,k)$ were obtained, where the $k$-qudit acts as the pivotal qudit.  
 Linear superposition of states was effected at this position, while the state of qudits at the other positions were transformed to another state by certain exchange operators.  Multi-qudit states of varying degree of entanglement can be generated by various combinations of $B_{0l}(n,k)$'s with different $l$ and $k$.  Particularly, we showed that the generalized GHZ states could be obtained directly without resorting to further local unitary transformations as in the Yang-Baxter approach.
 
For definiteness we have presented our construction of the braid generators and the braid quantum gates using
 $\ket{0}$ as the pivotal qudit.  This is not necessary. In fact any qudit, say $\ket{q}, (q=1,2,\ldots, D-1)$, can serve as the pivotal qudit.  To obtain the corresponding braid generators and the braid quantum gates, we need only to exchange $\ket{0}$ and $ \ket{q}$ in the expressions presented in this work.  For instance, the braid quantum gate $B_{0l}(n,k)$ in Eq.\,(\ref{B0l}) becomes
\begin{eqnarray}
B_{ql}(n,k)\ &=& \otimes_{j=1}^{k-1}I
\otimes
 \left( d_l a_l^2 \ket{q}\bra{q}+ \left( d_l b_l^2 + A_l^{-2}\right)\ket{l}\bra{l}+ \sum_{m\neq q,l}\, A_l^{-2} \ket{m}\bra{m}\right)
\otimes_{j=k+1}^n\, I\\
&+& d_l a_l b_l\otimes_{j=1}^{k-1} \lambda_j^{(l)} \otimes
 \left( e^{i\varphi_l} \ket{l}\bra{q} - e^{-i\varphi_l}A_l^4 \ket{q}\bra{l}\right) \otimes_{j=k+1}^n
\lambda_j^{(l)}\,, ~~~~   l \neq q\n
\end{eqnarray}
One then applies $B_{ql}(n,k)$ on the separable state $\ket{s_1 s_2 \cdots , q, \cdots s_n}$ to generate other linear superimposed states.

We have also pointed out an interesting observation that for a general multi-qudit state there exists a unitary braid quantum gate based on the Temperley-Lieb algebra that connects the multi-qudit with one of its component basis state, provided the coefficient of the component state is such that the square of its norm is no less than $1/4$.

%%%%%%%%

\section*{Acknowledgments}

The present research is partially supported by the Ministry of Science and Technology (MoST)
of the Republic of China under Grants  102-2112-M-032-003-MY3 and 105-2918-I-032-001 (CLH), 
and the Grant-in-Aid for Scientific Research No. 15K05204 (TD). 
This work was done during CLH's visit at the Ochanomizu University.  
He would like to thank  the members of the Department of Physics, particularly  E. Uehara, E. Nozawa, C. Matsuyama, and N. Oshima,  for the hospitality extended to him.

%%%%%%%%%%

%----------------------
\end{document}